\documentclass[acmsmall]{acmart}

\AtBeginDocument{%
  \providecommand\BibTeX{{%
    \normalfont B\kern-0.5em{\scshape i\kern-0.25em b}\kern-0.8em\TeX}}}

\newcommand{\soutc}[1]{\out{}}

\usepackage{wrapfig}
\usepackage{caption}
\usepackage{subcaption}
\usepackage{csquotes}
\usepackage{makecell}
\usepackage{longtable}

\usepackage{setspace}

\usepackage{ulem}

\usepackage{cancel}

\usepackage{color, colortbl}
\usepackage{wrapfig}

\setcopyright{none}
\renewcommand\footnotetextcopyrightpermission[1]{} 

\definecolor{Gray}{gray}{0.9}

\definecolor{white}{RGB}{234, 239, 249}
\definecolor{LightGreen}{RGB}{238, 238, 238}
\definecolor{LightOrange}{RGB}{221, 221, 221}
\definecolor{LightPink}{RGB}{204, 204, 204}

\begin{document}

\title{Quantifying the Invisible Labor in Crowd Work}



\author{Carlos Toxtli}
\affiliation{%
  \institution{Northeastern University}
  \city{Boston}
  \country{United States}}
\email{carlos.toxtli@mail.wvu.edu}

\author{Siddharth Suri}
\affiliation{%
  \institution{Microsoft Research}
  \city{Redmond}
  \country{United States}}
\email{suri@microsoft.com}

\author{Saiph Savage}
\affiliation{%
  \institution{Northeastern University \& Universidad Nacional Autonoma de Mexico (UNAM)}}
\email{s.savage@northeastern.edu}





\renewcommand{\shortauthors}{Carlos Toxtli, Siddharth Suri, \& Saiph Savage}

\begin{abstract}
Crowdsourcing markets provide workers with a centralized place to find paid work. What may not be obvious at first glance is that, in addition to the work they do for pay, crowd workers also have to shoulder a variety of unpaid invisible labor in these markets, which ultimately reduces workers' hourly wages. Invisible labor includes finding good tasks, messaging requesters, or managing payments. However, we currently know little about how much time crowd workers actually spend on invisible labor or how much it costs them economically. To ensure a fair and equitable future for crowd work, we need to be certain that workers are being paid fairly for {\it all} of the work they do. In this paper, we conduct a field study to quantify the invisible labor in crowd work.  We build a plugin to record the amount of time that 100 workers on Amazon Mechanical Turk dedicate to invisible labor while completing 40,903 tasks. If we ignore the time workers spent on invisible labor, workers' median hourly wage was \$3.76.  
But, we estimated that crowd workers in our study spent 33\% of their time daily on invisible labor, dropping their median hourly wage to \$2.83. We found that the invisible labor differentially impacts workers depending on their skill level and workers' demographics. The invisible labor category that took the most time and that was also the most common revolved around workers having to manage their payments. The second most  time-consuming invisible labor category involved hyper-vigilance, where workers vigilantly watched over requesters' profiles for newly posted work or vigilantly searched for labor. We hope that through our paper, the invisible labor in crowdsourcing becomes more visible, and our results help to reveal the larger implications of the continuing invisibility of labor in crowdsourcing.

\end{abstract}

\settopmatter{printfolios=true}
\maketitle

\section{Introduction}
Crowdsourcing markets, and their APIs, can help absorb some of the costs associated with crowd work \cite{gray2019ghost,lustig2020stuck}. From the requesters' perspective, these platforms provide an always-available pool of workers and an easy-to-use payment API to contract workers and start getting work done   \cite{huang201710,bernstein2011crowds,lasecki2013warping}.  From the workers' perspective, these markets provide a central place to find work and offer them the flexibility of working from wherever they desire \cite{alkhatib2017examining,martin2014being}. 

However, recent research has identified that some of these costs do not actually get absorbed by the crowdsourcing platform, but rather, they are passed onto the workers in the form of invisible labor \cite{gray2019ghost,crain2016invisible}. 
Invisible labor is defined as {\it ``unpaid activities that occur within the context of paid employment that workers perform in response to requirements (either implicit or explicit) from employers and that are crucial for workers to generate income, to obtain or retain their jobs, and to further their careers, yet are often overlooked, ignored, and/or devalued by employers, consumers, workers, and ultimately the legal system itself \cite{crain2016invisible}.}'' 

Invisible labor is also present in crowd work and it includes activities such as: the unpaid time workers have to invest in finding work, figuring out on their own how to complete the job at hand, or managing their payments \cite{kaplan2018striving,gray2019ghost}. The problem is that crowd workers are forced to engage in these unpaid activities just to be able to complete the labor for which they are paid \cite{qiu2020improving,sutherland2018sharing}. If we are aiming to create a future where crowd work is fair and equitable to workers, we need to ensure that workers receive a fair wage for {\it all} of the labor they do, whether it is the actual tasks for which they get paid, or the unpaid invisible work they do above and beyond that work. 

The central question this work addresses is how much time do workers actually spend on invisible work, and how does this affect their overall hourly wages? This is an important question not only to ensure that workers receive a fair wage now but also to ensure that workers receive a fair wage in the future.  Notice that our research is addressing a critical problem because a common use case for crowd work is to train machine learning algorithms, or to provide a human-in-the-loop approach when A.I. fails \cite{su2012crowdsourcing,shadbolt2013towards,cheng2015flock,gao2021human} . Since we are in the midst of an ``A.I. revolution,'' it is plausible that we will see dramatic growth in the use of crowd labor \cite{savage2020solidarity,jarrahi2018artificial, jarrahi2021flexible,carmel2020five,garcia2019impact}. In addition, post-COVID-19, there will likely be a large increase in people who need to work from home, whether that is for safety reasons or because of the massive number of worldwide layoffs \cite{spurk2020flexible,duke2020effects,fairWOrkOxford}. Measuring invisible labor in crowd work will only grow in importance going forward.

To start to quantify the invisible labor in crowd work, we develop a web plugin\footnote{https://github.com/anonym-research/invisible-labor} that allows us to detect when a worker is performing invisible labor and quantitatively measure the amount of time the worker spends on such efforts on Amazon Mechanical Turk (MTurk), one of the most popular crowdsourcing markets \cite{kuek2015global}.  We conduct a field study with our plugin to measure in the wild the amount of time crowd workers invest in invisible labor, and contrast with the amount of time workers spent on paid labor. Armed with our plugin, we had 100 crowd workers use our plugin for one week. Workers completed 40,903 human intelligence tasks (HITs). Through our plugin, we uncovered that crowd workers spent 33\% of their time on MTurk doing unpaid work. Relatively similar to prior work, we found that workers' median hourly wage considering only paid labor was  \$3.76 \cite{hara2018data}. But, if we consider the time workers spent on invisible labor, we calculated that workers' median hourly wage dropped to \$2.83. We also found that the amount of time that workers dedicated to invisible labor varied across workers' skill level and demographics. We found that master workers spent 23\% less time on invisible work than regular workers. We also observed that the time spent in invisible work appears to be heavily correlated with  demographic factors. 

The invisible labor in which crowd workers spent the greatest portion of their time revolved around payments. In particular, this most time consuming activity involved doing tasks for which workers were not paid because they experienced a ``time out'' (and hence they did not receive any payment for any of the labor they conducted for the task). Workers spent a median of 4.5 minutes daily on this activity. Overall, invisible labor around ``payments'' was the most time-consuming for workers; it was also among the most common.  In fact, 97\% of the workers in our study practiced invisible labor around visiting the earnings section on their workers' dashboard (perhaps to ensure they had gotten paid fairly \cite{whiting2019fair}). The second most time-consuming category of invisible labor involved hyper-vigilance where workers were ``on-call'' vigilantly watching over requesters' profiles ready to do, at all hours of the day, the labor that certain requesters posted, as well as vigilantly searching for work on Amazon Mechanical Turk \cite{williams2019perpetual,difallah2015dynamics,gray2019ghost}. Understanding invisible labor is key to creating positive change in crowd work \cite{crain2016invisible}; however, it has remained so far understudied. Bettering our understanding of invisible labor will allow us to design fairer crowdsourcing markets. 

\section{Related Work}

Our research builds on two main pieces of literature: (1) research on invisible labor, especially within digital labor markets \cite{crain2016invisible}; and (2) methodologies to quantitatively measure the time crowd workers spend completing paid labor on MTurk \cite{saito2019turkscanner,savage2020becoming}, i.e., HITs.

\subsection{Invisible Work}
The literature has traditionally characterized invisible labor as work that is ``economically devalued through cultural, legal, and/or spatial dynamics'' \cite{hatton2017mechanisms}. Under this definition, invisible labor is usually done in private rather than public \cite{daniels1987invisible,gouthro2009neoliberalism}. Usually, housework is one of the most commonly cited examples of invisible labor \cite{eichler2007household,ciciolla2019invisible}, and it involves both psychical labor and mental labor (e.g., planning what types of household chores should be done and in what manner.) 

In their book Invisible Labor \cite{crain2016invisible}, Crain et al. describe the concept of invisible work as the “{\it activities that occur within the context of paid employment that workers perform in response to requirements (either implicit or explicit) from employers}.” They explain how this concept has existed in different offline settings before, but nowadays, technology has enabled a large part of all invisible labor.  In particular, a number of technology companies are passing several aspects of digitization labor to consumers and workers, e.g., consumers are expected to install all the required Internet infrastructure at their homes. This labor is typically presented as something that is mundane, flexible, and part of the ``do-it-yourself'' culture \cite{glucksmann2016completing}. However, this dynamic also reduces the meaning of this type of labor, making it invisible, and something for which people are not paid. 
\subsection{Invisible Labor in Crowd Work}
Gray and Suri \cite{gray2019ghost} explored these concepts of invisible labor with a particular angle toward crowd workers. Through in-depth interviews with crowd workers, their book ``Ghost Work'' unveils the current conditions to which crowd workers are exposed and explains how companies have placed on the shoulders of workers a great portion of the invisible labor that companies themselves would traditionally do. The book also went a step further and started to describe the different types of invisible labor present in crowd work. 
Within this setting of describing invisible labor in crowd work, it is important to consider that crowd work does not emerge only from the requesters' side of the market; it is also something that crowdsourcing platforms, such as Amazon Mechanical Turk facilitate \cite{gray2019ghost} and could, with different design choices, help alleviate. For instance, crowdsourcing platforms could help match workers to tasks to reduce search time. Also, they could potentially pay workers for the time they spend searching for tasks or reading messages from requesters, which is something that companies have traditionally covered \cite{gray2019ghost,de2015rise}. 

Furthermore, when thinking about invisible labor in crowd work, we have to consider that much of the labor that crowd workers complete is fed into machine learning models that power the A.I. industry \cite{kittur2013future,su2012crowdsourcing}. For instance, crowd workers might label content so that Facebook's News Feed algorithm will not recommend posts that are filled with hate speech or pedophilia \cite{dang2018but,gillespie2018custodians}.  Crowd workers might also transcribe audio to help Amazon's Alexa better understand the user \cite{bigham2017deaf}. Given that most end-users are unaware that there are humans helping to power the A.I. services they access \cite{gray2019ghost}, the work done by workers and their possible unfair labor conditions, are hidden from sight. Notice that here the invisibility of crowd work is again not just due to requesters and their HIT design choices, but rather it is an issue within the A.I. industry as a whole.

In this particular research, we focus on measuring the different categories of invisible work that the book of Ghost Work identified that exist within the context of crowd work \cite{gray2019ghost}. We believe that by quantifying the different costs that invisible work has in this setting, we can design better solutions to improve crowd workers' conditions. Notice that invisible labor in crowd work includes activities that go unnoticed while doing paid work, such as finding HITs and communicating with requesters to resolve conflicts \cite{gadiraju2017modus,sannon2019privacy,han2019all,hara2018data}. Invisible labor in crowd work has recently gained more attention because it has become clearer that the independent nature of crowd work has led workers to now have to assume invisible labor that was traditionally taken by companies and employers \cite{rzeszotarski2011instrumenting,de2015rise}.
In this work, we present computational mechanisms for quantifying for the first time the invisible labor that exists in crowd work and bring much needed light to a critical topic. 

\subsection{Quantifying Working Time}
Saito et al. \cite{saito2019turkscanner} studied different ways to measure the time crowd workers spent completing HITs using their system called TurkScanner. They found that through web plugins, they could quantify how much time workers spent completing HITs. We built upon their methodology to properly measure the working time on HITs and expand their method to now also measure time spent in unpaid work. Hara et al. \cite{hara2018data} also used plugins to measure the wages of MTurk workers. Ignoring unpaid work, they estimated an average hourly wage of \$3.18, which roughly coincides with the average hourly estimate of \$3.76 that our study uncovered when we also ignore unpaid work. This shows that our measurement instrument is roughly calibrated to theirs (we likely had higher hourly wages because we considered more conservative measurements). We build on their work and provide a more detailed accounting and classification of the unpaid work that workers endure.

\section{Methods}
The goal of our IRB-approved field study is to measure and contrast the time that crowd workers spent on invisible labor and the time they spent on regular paid labor (i.e., completing HITs.) Since this data is not part of the official MTurk API, and prior work has not been able to measure invisible labor at the level of detail in which we were interested, we build computational mechanisms to measure these variables. Armed with these computational mechanisms, we conduct a field study to investigate in the wild how much time workers on MTurk dedicate to invisible labor. In the following, we describe how we measured these activities through the computational mechanisms that we designed and detail how we conducted our field study. It is important to highlight that our computational methods for measuring invisible labor focus on measuring invisible labor in a conservative manner. We consider it is best to err in underestimating the amount of time that workers spend in invisible labor than to overestimate. We make this design decision because quantifying invisible labor can potentially call attention to the structural issues surrounding crowdsourcing markets and the conditions they provide workers. Operating in a conservative manner helps us to avoid being labeled as ``exaggerated activists'' and allows us to present the study in a scientific, objective way. This approach helps us to bring much-needed attention to understanding invisible labor in crowd work. As we will see, even with erring on the side of underestimating invisible labor, it is still a sizeable overhead for the workers. 

\subsection{Computational Mechanisms to Measure Invisible \& Paid Labor}

For our study, we need computational mechanisms for: (1) detecting when a worker is doing invisible labor or when she is doing paid work; and (2) measuring how much time a worker invests in each of these two activities. To address these two points, we created a Chrome Extension (plugin). 

\subsubsection{Methods for Quantifying  Paid Labor}
Our plugin builds on prior research that was able to detect and measure with plugins when a crowd worker was completing a HIT, the amount of time the worker invested in completing the HIT, and the daily earnings that workers made from the HITs (notice that this value is important as it can help us to quantify the monetary costs of invisible labor)\cite{saito2019turkscanner}. In particular, building on prior work, we developed a plugin that can: 1) automatically record the exact times when a worker accepts a HIT and when she finishes and submits the HIT; 2) track when a tab about a HIT is in focus and automatically record the time period in which the worker is active on the HIT page tab by checking whether there were any type of interactions from the worker (e.g., mouse movements, typing) under a given time window; and also 3) measure the daily income that each worker makes from these HITs by querying the information from their workers' dashboard on MTurk. In summary, as a starting point, we developed our own plugin that mimics prior work and quantifies the amount of time that a given worker dedicates to completing HITs and the earnings that the worker is making. 

\subsubsection{Methods for Quantifying Invisible Labor}
Next, we expand the plugin to now provide new functionality through which we can also track and measure the time workers spend on invisible labor. Notice that we focus on quantifying invisible labor in a conservative manner, which means that we prefer to err on the side of under measuring the invisible labor (we took this methodological decision based on the reasoning stated above). Our conservative approach to the measurement of invisible labor comes in especially when we consider  cases where there is disagreement in the literature on whether an activity is invisible work or not \cite{gillier2018effects,rzeszotarski2013inserting}. In such cases, we prefer not to label the activity as invisible labor. We prefer to underreport so that the invisible labor we measure will be at least as large as we quantify here, if not larger. Some of the discussions around what is and what is not invisible labor especially arise for the activities of ``reading instructions of the HITs'', and ``taking breaks'' \cite{gray2019ghost,10.1145/2499149.2499168,gadiraju2017clarity,gillier2018effects,lasecki2015effects,10.1145/2740908.2744109}. Gray and Suri \cite{gray2019ghost} label ``reading instructions'' and ``taking breaks'' as examples of invisible labor activities. However, we decided not to categorize these activities as invisible labor because there is research that considers these two activities as part of paid work \cite{10.1145/2499149.2499168,gadiraju2017clarity,gillier2018effects,lasecki2015effects,10.1145/2740908.2744109}. Now, given that workers are not actually paid for either of these two activities, we designed our computational methods to detect when workers take breaks or read instructions; but, we do not count these activities as either paid nor invisible labor. It is important to highlight that because workers are not paid any wages for reading instructions, it is incorrect to categorize the work as being paid.

Our plugin, therefore, in addition to what prior work had already developed, provides now the novelty of being able to detect and quantify all other activities that workers do aside from completing HITs. For this purpose, we developed new computational mechanisms to detect when a worker is visiting other parts of the MTurk platform that are different from the HIT page tab\footnote{https://worker.mturk.com/} (e.g., perhaps the worker entered the MTurk page to search for HITs\footnote{https://worker.mturk.com/?filters} or the worker entered the MTurk page for sending messages to requesters\footnote{https://worker.mturk.com/contact\_requester}). Our plugin tracks the exact time when a worker enters one of these other MTurk domain pages and then scrapes and parses the HTML of the page to understand how the worker interacted with the page and identifies the intervals of time in which the worker is active on these other pages. We consider a worker to be active on a page when the worker has the page in focus and does any type of user interaction on that page, e.g., mouse movements, scrolls, clicks, keyboard typing. Notice that we do not track what a worker does on these pages (e.g., we do not track what they type). We simply detect that they are active on a particular MTurk page. To accomplish all of this, we developed two new components into our plugin: a page crawler and a time-driven background process that detects the different browser events that happen on MTurk (e.g., that the worker visited another page on MTurk, or that she started typing, or began a new HIT). The page crawler detects the current MTurk domain page that the worker is on, as well as the status of the page (e.g., that the page is loaded, active, inactive, or closed). The background process focuses on detecting the HITs that the worker is currently doing and identifying which she has finished. In order to accomplish this, the background process polls workers' task queues on MTurk every 30 seconds. From the task queue, the background process obtains the metadata and status of all the HITs the worker has accepted to do. Notice that the page crawler is the primary element that we use to detect whether the worker is completing paid labor or invisible labor. The background process helps our plugin to be able to better detect when the worker is completing HITs (some of them reside outside the MTurk platform) and also when the worker is multi-tasking (doing multiple HITs at the same time.) Through this, we create a plugin that automatically detects when a worker is doing invisible or paid labor and the amount of time the worker invested in each of these two activities. Our plugin is available here: \url{https://github.com/anonym-research/invisible-labor}.

\subsubsection{Quantifying Types of Invisible Labor.}
We were not only interested in detecting whether or not a worker was doing invisible work;  we also wanted to understand what type of invisible labor was the most taxing and contextualize our results with prior interview work that started to document the invisible labor that workers perceived by conducting interview studies with them \cite{gray2019ghost}. In the following, we present the different types of invisible labor we consider (i.e., broad categories) and how we detected their related individual activities. The categories and activities we study are based on prior interview research that studied invisible labor \cite{whiting2020digi,gray2019ghost}. Note that for most cases, we detect that a worker started a new activity when they loaded, focused, or changed their browser tab to a page on MTurk related to that particular type of invisible labor (below, we mention which pages relate to specific invisible labor activities). Similarly, our plugin considers that a worker paused or finished an activity when the worker changed to another tab, unloaded, blurred, or closed the MTurk page related to that particular activity. The categories and activities we consider are:\\

\textbf{(a) Category: Hypervigilance.}  This category involves workers spending time in: (1) identifying good work, e.g., {\it ``wading and sorting through spam or suspicious offers for at-home-work.}''\cite{deng2016individuals}; and (2) being ``on-call,'' ready to do HITs for requesters at any time. Invisible activities include:

\begin{itemize}
\item{{\it Watching over requesters' profile}}: Notice that this activity relates to Hypervigilance because workers are visiting requesters' profiles to be ready to do any HIT that requesters post. In other words, workers are ``on-call.'' To detect this activity, our page crawler detects when a worker is on a requester's profile page.
\item{{\it Searching for general HITs (unfiltered)}}: To detect this activity, our page crawler identifies that a worker is on the main page where HITs are posted.
\item{{\it Searching for filtered HITs}}: Our page crawler detects when the search URL for the main page of HITs has a query in it to filter HIT results. This activity relates to hypervigilance as it involves ``wading and sorting'' through HITs.
\item{{\it Managing their queued HITs}}: this activity relates to Hypervigilance as it involves workers filtering out fraudulent HITs and focusing on HITs from specific requesters (i.e., being ``on-call''). To detect this, our crawler identifies when a worker is visiting her tasks queue.
\item{{\it Checking their own qualifications}}: This activity relates to Hypervigilance as prior work has identified that workers watch over their own qualifications to vigilantly identify whether they could now access certain HITs and thus more effectively find and access quality labor \cite{gray2019ghost}. In this case, our crawler detects when the worker is viewing her earned qualifications.
\end{itemize}

\textbf{(b) Category: Lack of Guidance.} Crowd workers are generally left on their own to figure out how to do jobs as fast and accurately as possible \cite{margaryan2016understanding}. Activities related involve:
\begin{itemize}

\item{{\it Starting HITs but then returning them}}: This activity relates to ``Lack of Guidance'' as it usually occurs because workers believe the HITs will be different than what they actually end up being \cite{mcinnis2016taking,gol2018take} (e.g., less complex or of another nature.) The lack of guidance leads workers to have to return HITs they already started. In this case, our crawler detects when workers click the return HIT button on MTurk.
\item{{\it Sending messages}}: Workers send messages to requesters to ask them questions about a HIT and better understand what the requester wants. To detect this type of invisible labor, our crawler detects when a worker opens MTurk's messaging form to send a message.
\item{{\it Reading HIT information}}: Page crawler detects when a worker clicks the ``More Info'' option while previewing or working on a HIT. Notice that this activity is different from reading HIT instructions, as reading HIT information helps workers get a preview of what a HIT is about. It is an activity that workers have to do in order to obtain guidance.
\item{{\it Previewing HITs}}: Page crawler detects when the page of a HIT is open in preview mode.  Notice that here we could potentially say that workers are previewing HITs in order to ``vigilantly'' find tasks from certain requesters (and hence this activity could be labeled as being from the category of Hypervigilance). However, the search filtering option allows workers to do that more effectively, and that is also not the main purpose of the preview \cite{kittur2008crowdsourcing}. We, therefore, decided to categorize this activity as Lack of Guidance. Additionally, prior work has labeled this activity  as related to guidance \cite{wu2017confusing,manam2018wingit}.
\item{{\it Reading platform help}}: Page crawler detects when workers are in MTurk support sections.
\end{itemize}

Notice that within this category, we could have considered the activity of reading instructions as part of the invisible labor that a worker has to do related to the lack of guidance. However, as mentioned before, we opted to just detect the activity but not label it as invisible nor paid labor. To detect the activity of ``reading instructions,'' the page crawler detects the time that passes from when a worker accepted a HIT until the worker has her first interaction with the HIT (e.g., she presses a key, or she opens another tab related to the HIT, etc.) We assume that this time-lapse corresponds to when the worker is reading instructions.

\textbf{(c) Category: Payments.} In crowd work, even after workers have vigilantly identified legitimate labor and they have been able to figure out how to complete the work, they still run the risk that they will not get paid for their efforts. The broad category of ``Payments'' encompasses the invisible labor that workers do to ensure payment and also instances where they worked on HITs but were not paid. This category of invisible labor includes:
\begin{itemize}
\item{{\it Visiting their worker's dashboard}}: workers visit their dashboard to oversee if requesters have paid them and ensure they made a certain amount of daily income. To detect this activity, our crawler identifies when workers are visiting their general MTurk dashboard.
\item{{\it Doing HITs that eventually timeout}}: Some HITs have an expiration time on them. If workers take longer to complete the HIT than the allowed expiration time, the HIT times out. In these cases, workers are not paid for any of the labor they have done on the HIT, and thus we consider this activity within the broader category of Payments. To detect these instances, the background process of our plugin identifies when a HIT has an end time equal to or higher to the HIT expiration time. Our plugin also checks in the worker's dashboard that the worker was never paid for those HITs. 
\item{{\it Viewing their earnings}}: Page crawler detects when workers are in earning sections on MTurk.
\end{itemize}

\textbf{(d) Category: General Logistics.} The last category we detect relates to MTurk logistics. We focus on the activities of logging into MTurk. Our crawler detects when workers log into MTurk.

\subsubsection{Detecting and Processing Multi-Tasking.}
When measuring the time workers spent in completing HITs, it is important to properly detect when workers are multi-tasking and properly measure and account for the time they spent doing so \cite{lascau2019monotasking}. In our study, we refer to multi-tasking as when a worker accepts multiple HITs or batches of HITs around the same time and then starts completing these multiple HITs. The background process of our plugin checks the workers' tasks queue to detect workers completing HITs via multi-tasking. To account for this time, we adopt an approach similar to prior work \cite{saito2019turkscanner}. A common feature of working in this manner is that the HITs are chained in succession. This means that the start and end times may overlap with one or more HITs in the batch. 
Also, similar to prior work \cite{saito2019turkscanner}, our study does not consider batch HITs that take more than one day to be completed (0.6\% 
of our sample). We filtered out all the multi-day batches and HITs since these imply computing the effective working schedule of each worker. 
\subsection{Field Study}
The purpose of our field study was to have workers use our plugin and through it measure in the wild the amount of time workers dedicate daily to invisible labor. Armed with our plugin that could detect and measure the amount of time workers dedicated to different types of invisible labor, as well as the time they dedicated to paid labor, we conducted a field study that lasted a week. Note that we included weekends in our analysis as MTurk presents itself as a platform that offers workers the flexibility to work whenever workers want (weekends included). Similar to prior work \cite{hara2019worker,savage2020becoming}, we did not see changes in the days workers completed tasks. 
\subsubsection{Field Study Logistics.}   We recruited workers from 
MTurk by posting a HIT inviting workers to our study. We also used mailing lists of Turkers (workers on MTurk) who had participated previously in studies with us. For our study, we first surveyed participants on their perceptions of how much time they estimated that they spent on invisible labor. We asked workers to report how much time they felt they invested on MTurk: searching for work; looking over their worker dashboard; sending messages to requesters; and doing HITs that eventually timeout. 
This helped us understand workers' prior beliefs and awareness of invisible labor and how much time they believed they spent on it. We also asked workers about how COVID-19 had affected them (none of our participants expressed any work disruptions). Our initial survey also asked workers about their basic demographic information, such as current location, gender, disabilities, etc. Overall, we based our survey on prior work \cite{hara2019worker,flores2020challenges}. 

After the initial survey, we asked participants to: (1) install and use our plugin for a week; (2) work on MTurk as normal; (3) visit the plugin dashboard, which showed to each worker graphs of how much time the plugin detected that they invested in different MTurk activities for a given day. At the end of the field study, workers completed a short survey evaluating the accuracy of the plugin in detecting and measuring the amount of time they spent on different activities on MTurk. In general, workers in our study stated that they felt that our plugin was able to adequately track the time they spent daily on MTurk completing HITs and doing different invisible labor activities (the median score for the plugin's accuracy was 4 on a 5 point Likert scale). We paid each participant \$10 USD for taking part in our study. Notice that this accounts for the US federal minimum wage (\$7.25/hour) as our initial survey took 5-8 minutes to complete, the installation of our plugin took less than 4 minutes, and the end survey we gave participants took 5-8 minutes.

\section{Results}

\begin{table}[t]
\begin{center}
\begin{tabular}{c c}
 {\bf \tiny Description of the Statistics} & {\bf \tiny Value}\\ 
 \hline
 \rowcolor{white}
  {\tiny Total number of workers in our study} &{\tiny 100 
  }\\
 \rowcolor{white}
  {\tiny Total number of HITs workers did in a week} & {\tiny 40,903 
  }\\
  \rowcolor{white}
  {\tiny Minimum number of HITs a worker did in a week} &{\tiny 1 
  }\\
  \rowcolor{white}
  {\tiny Median number of HITs a worker did in a week}&{\tiny 185 
  }\\
  \rowcolor{white}
 {\tiny Maximum number of HITs a worker did in a week}&{\tiny 3,168 
 }\\
\rowcolor{white}
 {\tiny Minimum number of HITs a worker did per day} &{\tiny 0 
 }\\
\rowcolor{white}
 {\tiny Median number of HITs a worker did per day}&{\tiny 30 
 }\\
\rowcolor{white}
  {\tiny Maximum number of HITs a worker did per day}&{\tiny 1,149 
  }\\
 \rowcolor{LightGreen}
 {\tiny Minimum time a worker invested in completing HITs per day}&{\tiny 0 
 min}\\
\rowcolor{LightGreen}
 {\tiny Median time a worker invested in completing HITs per day}&{\tiny 1:07 
 hrs}\\
\rowcolor{LightGreen}
 {\tiny Maximum time a worker invested in completing HITs per day}&{\tiny 7:36 
 hrs}\\
\rowcolor{LightGreen}
{\tiny Minimum time a worker invested in invisible labor per day}&{\tiny 0 
min}\\
\rowcolor{LightGreen}
{\tiny Median time a worker invested in invisible labor per day}&{\tiny 33 
min}\\
\rowcolor{LightGreen}
{\tiny Maximum time a worker invested in invisible labor per day}&{\tiny 5:31 
hrs}\\
\rowcolor{LightOrange}
{\tiny Minimum earnings made by a worker in a week} &{\tiny \$0.92 
}\\
\rowcolor{LightOrange}
 {\tiny Median earnings made by a worker in a week}&{\tiny \$55.39 
 }\\
\rowcolor{LightOrange}
{\tiny Maximum earnings made by a worker in a week}&{\tiny \$542.06 
}\\
\rowcolor{LightOrange}
 {\tiny Minimum earnings made by a worker per day} &{\tiny \$0.01 
 }\\
\rowcolor{LightOrange}
 {\tiny Median earnings made by a worker per day}&{\tiny \$8.07 
 }\\
\rowcolor{LightOrange}
 {\tiny Maximum earnings made by a worker per day}&{\tiny \$178.62 
 }\\
\rowcolor{LightOrange}
 {\tiny Median hourly wage with invisible labor}&{\tiny \$2.83 
 }\\
\rowcolor{LightOrange}
  {\tiny Median hourly wage without invisible labor}&{\tiny \$3.76 
  }\\
 \rowcolor{LightPink}
 {\tiny Percentage of workers who multi-task} &{\tiny 96\% 
 }\\
 \rowcolor{LightPink}
  {\tiny Minimum number of batches a worker did in multi-tasking}&{\tiny 1 
  }\\
 \rowcolor{LightPink}
 {\tiny Median number of batches a worker did in multi-tasking}&{\tiny 32 
 }\\
\rowcolor{LightPink}
 {\tiny Maximum number of batches a worker did in multi-tasking}&{\tiny 333 
 }\\
\rowcolor{LightPink}
 {\tiny Minimum number of HITs a worker did in a batch}&{\tiny 2 
 }\\
\rowcolor{LightPink}
 {\tiny Median number of HITs a worker did in a batch}&{\tiny 3 
 }\\
\rowcolor{LightPink}
 {\tiny Maximum number of HITs a worker did in a batch}&{\tiny 689 
 }\\
\end{tabular}
\caption[Table caption text]{Summary statistics of the workers in our study with regard to: HITs workers did, the time they invested in working, workers' earnings, and their multi-tasking information.}
\label{table:stats}
\end{center}
\end{table}

We had 100 MTurk workers install and use our plugin for a week. We allowed all types of workers to participate in our study. This resulted in us recruiting 21 ``master workers'' and 79 ``non-master'' workers. Note that we considered that a worker was a master worker if we detected that they had completed at least one HIT with master qualifications.

Table \ref{table:stats} presents the statistics of the workers in our study and their general labor patterns. We had 73 men and 27 women, who had a median age of 30 years old. 41 participants were from the United States, 45 from India, five from Brazil, three from Italy, and the remaining six from Venezuela, Spain, Mexico, United Kingdom, United States Virgin Islands, and Thailand.

\begin{figure}[t]
\centering
\includegraphics[width=1.0\textwidth]{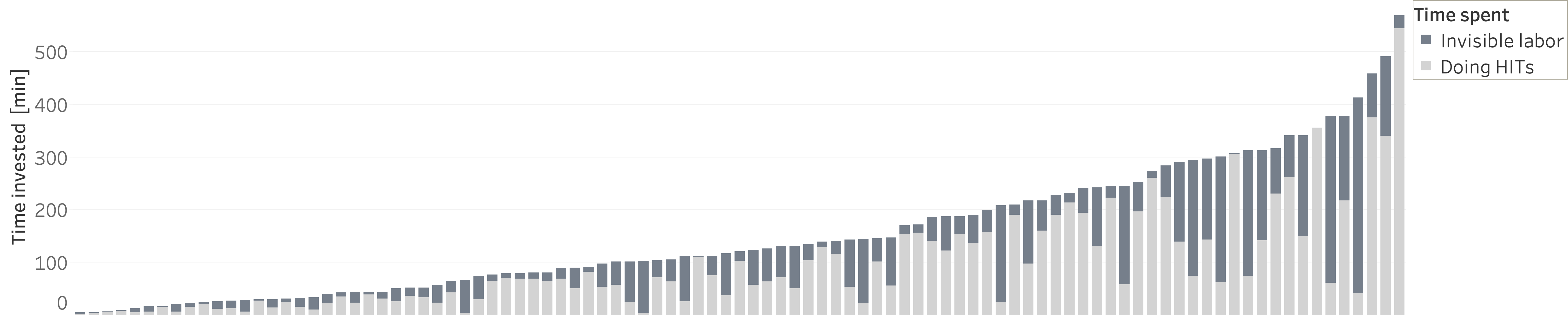}
\caption{Overview of the labor patterns of each worker in our study.
}
\label{fig:workerInvisibleAll}
\end{figure}

Through our plugin, we identified that workers did a median of 30 HITs each day. The median daily earnings of each worker were \$8.07 US dollars. Figure \ref{fig:workerInvisibleAll} presents the median amount of time that each worker invested in completing HITs during our one-week study. Each bar represents a worker, and the bars are sorted along the X-axis based on the median amount of time they worked daily on MTurk. The Y-axis shows the amount of time each worker dedicated to completing HITs or doing invisible labor. The light gray part of each bar shows how much time the worker spent doing HITs, and the dark gray part shows how much time they spend doing invisible labor. Observe that invisible labor occupied a substantial amount of workers' overall time. The median time that workers invested daily in completing HITs was 1 hour 7 minutes, and the median time that workers invested in invisible labor was an additional 33 minutes,  with some workers spending a maximum of 5 hours and 31 minutes daily. Notice that to calculate this value, we summed up all of the time that workers invested in the different invisible labor activities that our plugin detected. Workers spent a median of 33\% of their daily time on MTurk doing invisible labor.  

\begin{figure}
  \begin{center}
    \includegraphics[width=.6\linewidth]{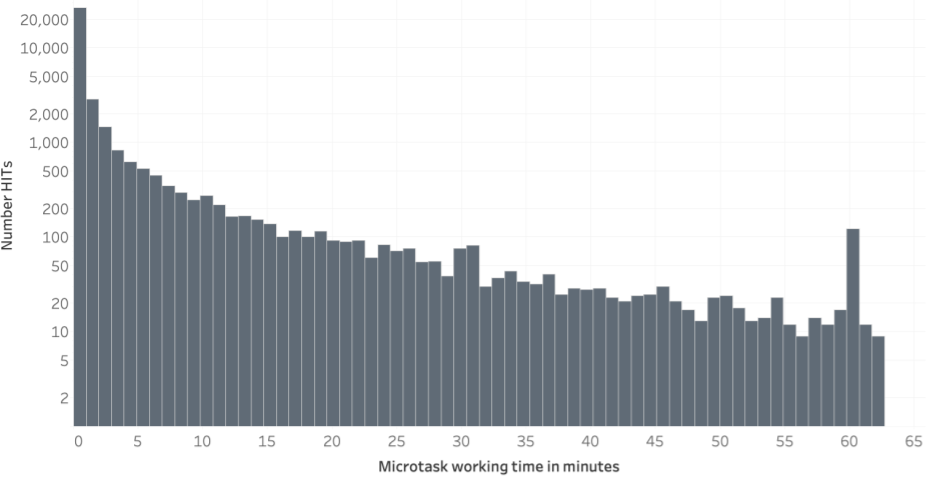}
  \end{center}
  \caption{Working time distribution of the HITs that crowd workers in our study completed.}
  \label{fig:workingTime}
  \vspace{-4pt}
\end{figure}

We also graphed a histogram of the amount of time that workers dedicated to completing HITs (see Fig. \ref{fig:workingTime}). This graph helps to calibrate whether our plugin is measuring paid labor adequately as we can compare our findings to prior work \cite{savage2020becoming}. Note that we used a log scale on the y-axis so that the distribution was easier to visualize. From here, we observe that similar to prior work \cite{saito2019turkscanner}, the distribution of the time that workers invested in completing HITs had a long tail that was heavily weighted towards shorter tasks, meaning workers usually did HITs that took under a minute.

Next, we were interested in studying whether there was a significant correlation between the time workers spent working and the time they spent conducting invisible labor (as this can help us to better understand the phenomena of invisible labor). For this purpose, we computed the Spearman's correlation and obtained 0.283 (p-value 0.004) for the time workers spent working and time doing invisible labor, and 0.517 (p-value 0.000) for the percentage of time working and time in invisible labor. Given these values, for both cases, we reject the null hypothesis that the samples are uncorrelated, i.e., we identified that there is correlation between the time workers' spent working and the time they spent completing invisible labor. Future work could thus study the type of paid labor that might minimize the amount of time a worker has to dedicate to invisible labor.

\begin{figure}
  \begin{center}
    \includegraphics[width=.3\linewidth]{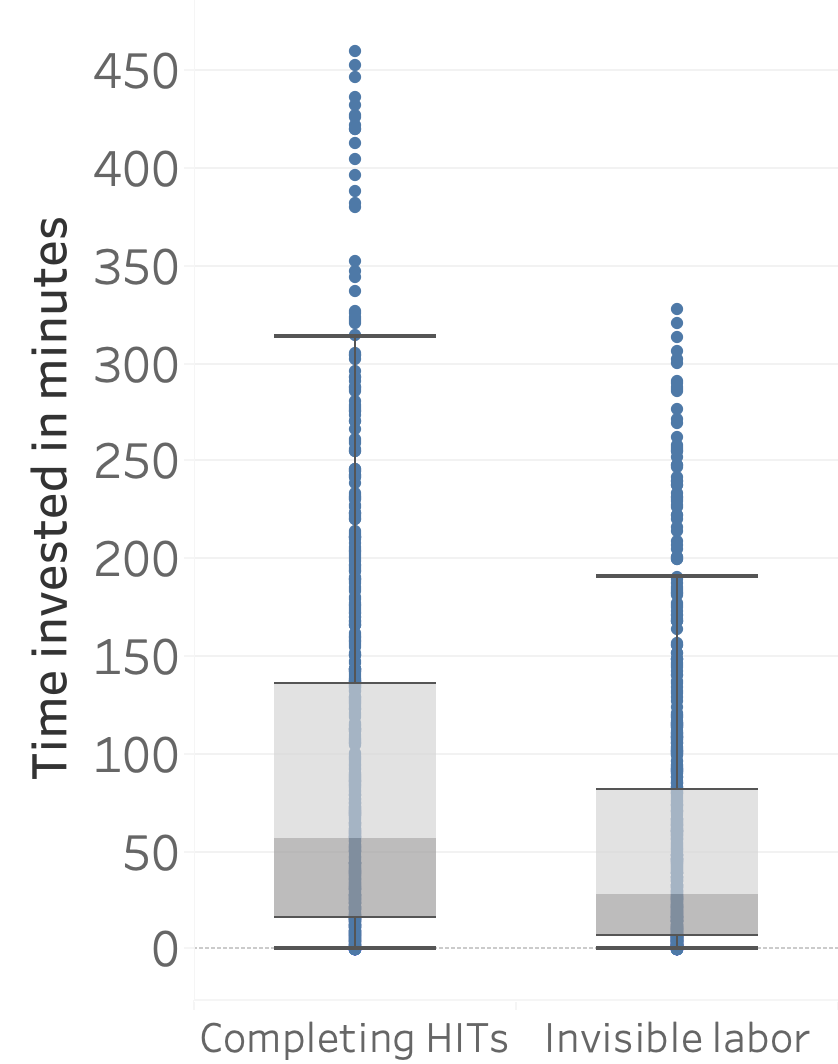}
  \end{center}
  \caption{Box plots showing the time that workers dedicate to completing HITs and doing invisible labor.}
  \label{fig:BoxworkingTime}
  \vspace{-4pt}
\end{figure}

\begin{figure}
  \begin{center}
    \includegraphics[width=.6\linewidth]{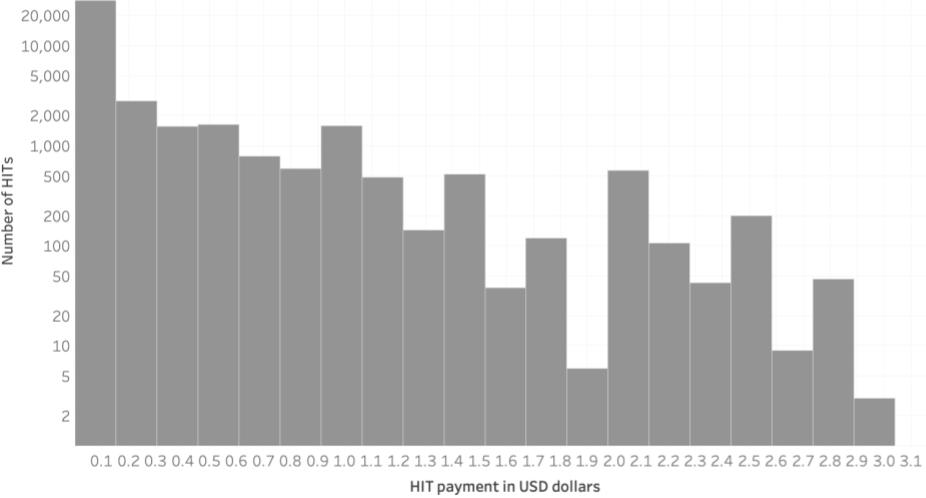}
  \end{center}
  \caption{Payment distribution of the HITs that crowd workers in our study completed.}
  \label{fig:tasksPayments}
  \vspace{-4pt}
\end{figure}

\subsubsection{Quantifying Invisible Labor and its Economic Costs.} We aimed here to understand the economic costs that invisible labor has on workers' wages. For this purpose, we first visualized in greater depth the median time workers spent daily in invisible labor and in completing HITs (see Fig. \ref{fig:BoxworkingTime}). We also aimed to understand the distribution of payments of the completed HITs (see Fig. \ref{fig:tasksPayments}). Armed with this information, we calculated the median hourly wage of workers. We used an approach similar to prior work \cite{savage2020becoming,hara2018data}. We first calculated the total hours a worker spent completing HITs on a given day D. We call this the worker's ${WorkingHour}_D$, and it is the sum of all the time series ($\text{Time}_{S}$), measured in hours, that the worker dedicated to doing HITs on day D within the time period $d$: 
\begin{equation}
    {WorkingHour}_D = \Sigma_{d\in D}^{} \text{Time}_{S,d}
\end{equation}
After this, we obtain the total {$Income_D$} the worker made on day $D$. We take this value from the rewards and bonuses logged on the worker's ``\text{Daily Income}'' on her MTurk dashboard. For worker $w$, her overall hourly wage for day $D$ is:
\begin{equation}
w_D = \frac{\text{{Income}}_D}{{{{WorkingHour}_D}}}.
\end{equation}
 With this, we calculate for each worker her hourly wage for each day of our study. We then use that information to calculate the median hourly wage of the 100 workers participating in our study. 
 
\begin{figure*}[t]
\centering
\includegraphics[width=.7\textwidth]{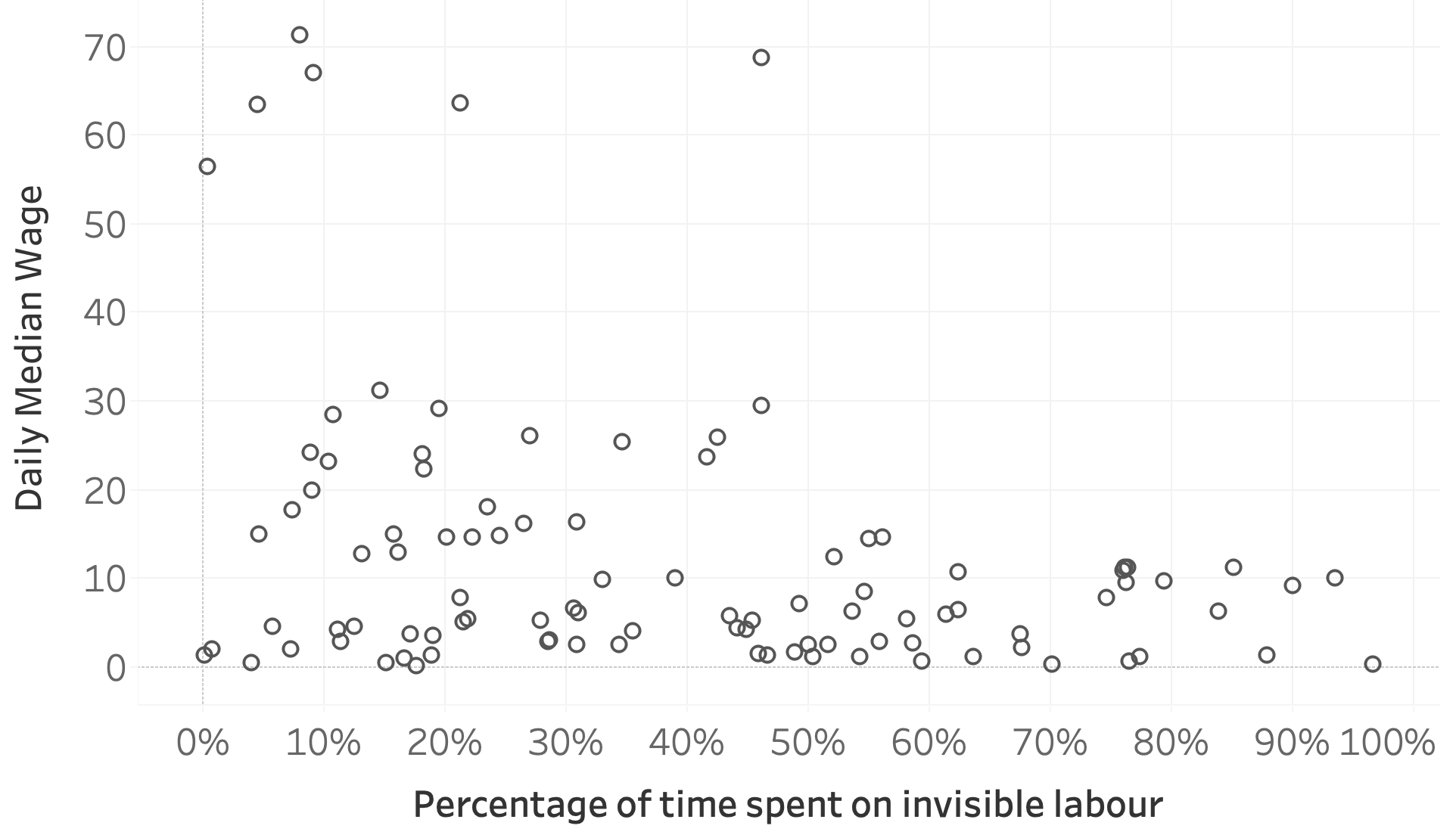}
\caption{Overview of the percentage of time each worker spent daily on invisible labor (X-axis) and their median daily wage (Y-axis).}
\label{fig:wagesInvisible}
\end{figure*}

 Excluding invisible labor, we calculated that workers earned a median hourly wage of \$3.76, which roughly coincides with prior work, which calculated \$3.18 \cite{hara2018data}. Notice that it is likely that we calculate a slightly higher salary because we utilize a slightly more conservative approach for our measurements, with the purpose of limiting the overreporting of invisible labor that workers do. Now, if we include invisible work into the calculation of the hourly wage, the median hourly wage of workers drops to \$2.83. 
 Next, we were interested in better understanding the dynamics around invisible labor and wages. Figure \ref{fig:wagesInvisible} presents a scatter plot where each point represents a worker. The X-axis represents the median percentage of time a worker invested in invisible labor daily, and the Y-axis the worker's median daily wage. From Fig. \ref{fig:wagesInvisible}, we observe that the highest-earning workers, in general, all invested less than 50\% of their time in invisible labor. Given this result, there might be value in exploring coaching systems that teach workers how to best manage their invisible labor to ensure high wages.
 
\subsubsection{Invisible Work for Different Segments of Workers}
\begin{table}[t]
\begin{center}
\begin{tabular}{c c c c c c}
 {\bf \tiny Workers' segment}& {\bf \tiny Working Time}& {\bf \tiny Invisible Work}& {\bf \tiny Unpaid rate}& {\bf \tiny Payment}& {\bf \tiny \# workers}\\ 
 \hline
 
 \rowcolor{white}
 {\tiny Masters Workers}& {\tiny 1hr 37min }& {\tiny 24min }& {\tiny 19.8\%}& {\tiny \$13.8 }& {\tiny 21 } \\
 \rowcolor{white}
 {\tiny Non-Masters Workers}& {\tiny 58min }& {\tiny 43min }& {\tiny 42.5\%}& {\tiny \$5.5 }& {\tiny 79 } \\
 
 {\tiny Workers based in United States (English Speaking)}& {\tiny 1hr 28min }& {\tiny 27min }& {\tiny 23.4\%}& {\tiny \$11.9 }& {\tiny 41 } \\
 {\tiny Workers based in India (English Speaking)}& {\tiny 42min }& {\tiny 35min }& {\tiny 45.4\%}& {\tiny \$4.0 }& {\tiny 45 } \\
 {\tiny Workers based in Brazil (non-English Speaking)}& {\tiny 18min }& {\tiny 1hr 15min }& {\tiny 80.6\%}& {\tiny \$1.9 }& {\tiny 5 } \\
 {\tiny Workers based in Italy (non-English Speaking)}& {\tiny 1hr 11min }& {\tiny 1hr 34min }& {\tiny 56.9\%}& {\tiny \$9.5 }& {\tiny 3 } \\
 
 \rowcolor{white}
 {\tiny Women}& {\tiny 1hr 02min }& {\tiny 42min }& {\tiny 40.3\%}& {\tiny \$8.2 }& {\tiny 27 } \\
 \rowcolor{white}
 {\tiny Men}& {\tiny 53min }& {\tiny 28min }& {\tiny 34.5\%}& {\tiny \$5.5 }& {\tiny 73 } \\
 
 {\tiny 18-24 years old}& {\tiny 26min }& {\tiny 33min }& {\tiny 55.9\%}& {\tiny \$3.3 }& {\tiny 9 } \\
 {\tiny 25-34 years old}& {\tiny 1hr 01min }& {\tiny 45min }& {\tiny 42.4\%}& {\tiny \$6.3 }& {\tiny 52 } \\
 {\tiny 35-44 years old}& {\tiny 59min }& {\tiny 22min }& {\tiny 27.1\%}& {\tiny \$5.9 }& {\tiny 22 } \\
 {\tiny 45-54 years old}& {\tiny 55min }& {\tiny 20min }& {\tiny 26.6\%}& {\tiny \$6.6 }& {\tiny 9 } \\
 {\tiny 55-64 years old }& {\tiny 1hr 12min }& {\tiny 36min }& {\tiny 33.3\%}& {\tiny \$9.8 }& {\tiny 6 } \\
 {\tiny 65-74 years old }& {\tiny 21min }& {\tiny 19min }& {\tiny 47.5\%}& {\tiny \$1.3 }& {\tiny 2 } \\
 
 \rowcolor{white}
 {\tiny No impairment declared}& {\tiny 54min }& {\tiny 31min }& {\tiny 36.4\%}& {\tiny \$5.7 }& {\tiny 93 } \\
 \rowcolor{white}
 {\tiny Mobility impairment}& {\tiny 1hr 19min }& {\tiny 19min }& {\tiny 19.3\%}& {\tiny \$15.9 }& {\tiny 5 } \\
 \rowcolor{white}
 {\tiny Mental disorder}& {\tiny 1hr 16min }& {\tiny 1hr }& {\tiny 44.1\%}& {\tiny \$9.2 }& {\tiny 2 } \\
 
 {\tiny Frequently multi-task}& {\tiny 1hr 04min }& {\tiny 28min }& {\tiny 30.4\%}& {\tiny \$6.7 }& {\tiny 42 } \\
 {\tiny Rarely multi-task}& {\tiny 48min }& {\tiny 32min }& {\tiny 40.0\%}& {\tiny \$5.4 }& {\tiny 58 } \\
 
 \rowcolor{white}
 {\tiny Use tools}& {\tiny 1hr 03min }& {\tiny 37min }& {\tiny 37.0\%}& {\tiny \$6.8 } & {\tiny 79 } \\
 \rowcolor{white}
 {\tiny Not use tools}& {\tiny 33min }& {\tiny 21min }& {\tiny 38.8\%}& {\tiny \$3.1 }& {\tiny 21 } \\
 
\hline 
\end{tabular}
\caption{Median of times and payments per segment. The unpaid rate shows the percentage of the total working time that is unpaid (invisible work). The payment amount represents the median daily payment.}
\label{table:segments}
\end{center}
\end{table}
In this section, we provide a breakdown of the different demographics of workers in our study (segments) and study the type of invisible labor they presented in their work practices. This analysis is important as research has started to showcase how workers' different demographics can impact how they approach work on MTurk \cite{hara2019worker,flores2020challenges,newlands2020crowdwork,simonshope}. We were thus interested in further studying and understanding this aspect, but now for invisible labor. In Table \ref{table:segments}, we present an overview of the amount of paid labor and invisible labor that different population segments conducted. Notice that in the table, we also calculate the ``unpaid percentage ratio,'' which denotes the percentage of the total working time that is unpaid. We calculated the unpaid percentage rate as follows, where $Invisible\_Labor\_Time$, is the median time workers in a particular segment spent on invisible labor, and $Paid\_Labor\_Time$ the median time workers in that segment spent on paid work. 

\begin{equation}
    {Unpaid\_Rate}= \frac{ {Invisible\_Labor\_Time}}{ {Invisible\_Labor\_Time}+{Paid\_Labor\_Time}}
\end{equation}

 Armed with these measurements, we next conducted statistical analysis to study whether there were significant differences between how invisible labor impacted the different segments of workers. First, over each worker segment we performed the Shapiro-Wilk test, which allows us to identify whether our distribution is normal or not. We found that for all the segments, the p-value was less than .05, so we rejected the null hypothesis (i.e., our distribution is not normal). Given that we do not have a normal distribution, we proceeded to use a non-parametric analysis of variance. We performed a Kruskal-Wallis H Test as a non-parametric alternative to the parametric one-way between-groups analysis of variance for independent groups. We found that there was a significant difference (p-value < 0.05) in the invisible labor time between the workers who were: Masters and Non-masters (p-value 0.00), male and female (p-value 0.04), tool users and non-tool users (p-value 0.01), from English speaking countries and non-English speaking countries (p-value 0.02). We did not find a significant difference among the following groups: workers without disabilities and workers with some disabilities (p-value 0.64); workers who do multi-tasking and workers who do no multi-tasking (p-value 0.32); workers in the U.S. and workers in India (p-value 0.07).

Next, we dug deeper into several of these results to better understand the dynamics behind invisible labor and workers' demographics. Table \ref{table:segments} shows that from the 100 Turkers who participated in our study, 21 of these were MTurk Masters, and 79 were not. We found that the median amount of time that master workers invested in completing HITs daily was 1 hour and 37 min and the median amount of time they invested in invisible work daily was just 24 minutes, as shown in Table \ref{table:segments}. Non-master workers worked slightly less time on HITs and spent more time on invisible labor than master workers. Non-master workers spent a median of 58 minutes daily completing HITs and a median of 43 minutes on invisible labor (almost double the time to what master workers invested.) Thus, workers with the Masters distinction spent more time working and less time doing invisible work than non-masters. Overall, a key takeaway from Table \ref{table:segments} is that Master workers perform 23 percentage points less invisible work than non-Masters workers (20\% vs. 43\%)  and earn a median of \$8.3 more a day. Naturally, 21 Masters is not a huge sample, so one should view this result as suggestive and follow up with future work to confirm. There are also a variety of explanations for this finding. It could be that the experience and know-how of the Masters workers help them minimize the amount of time they spent doing invisible work. Similarly, it could also be that Masters workers have more experience using tools. 86\% of our Masters worker participants reported using tools, while only 57\% of the non-masters workers reported tool use. Additionally, these workers might not be using these tools as effectively as the master workers. Prior work had identified that there are differences in how experts and non-experts use tools \cite{kaplan2018striving, savage2020becoming}. 

However, it is important to highlight that Table \ref{table:segments} does show that workers who used tools spent more time doing paid work (30 minutes more) and earned substantially higher wages (\$1.3 USD more daily, when measuring workers' median wages.) Notice that these results might be emerging because most tools focus on increasing the wages that workers receive for their paid labor \cite{kaplan2018striving}. But, given our results, we believe there is value in exploring mechanisms through which workers learn how to better navigate crowdsourcing markets to focus primarily on paid work.

Within this study of worker segments, we also studied the relationship between adopting particular strategies and invisible labor. Prior work has shown that experienced workers often use strategies to boost their performance \cite{han2020crowd,savage2020becoming}. This can include using different tools or multi-tasking. Our study identified that workers who completed HITs in batches did 9.6\% less invisible work than workers who did not (see Table \ref{table:segments}). The reasoning behind this finding is likely that within batches, the same type of tasks is continuously presented to workers (one after the other). Therefore, workers do not have to search for new tasks (thus reducing their invisible labor). Batch tasks are also usually similar, so workers do not have to spend time context switching \cite{lascau2019monotasking}.

\begin{table}[t]
\begin{center}
\begin{tabular}{c c c c c}
 {\bf \tiny Invisible Labor Activity}& {\bf \tiny Mean [min]}& {\bf \tiny Median [min]}& {\bf \tiny Std [min]}& {\bf \tiny \% workers}\\ 
 \hline
 \rowcolor{white}
 {\tiny Doing HITs that eventually timeout (Payments)}& {\tiny 32.3 
 }& {\tiny 4.5 }& {\tiny 1.5 }& {\tiny 37\% 
 }\\
 {\tiny Starting HITs but then return (Lack of Guidance)}& {\tiny 11.2 
 }& {\tiny 4.2 }& {\tiny 12.1 }& {\tiny 92\% 
 }\\
 \rowcolor{white}
 {\tiny Viewing their worker's dashboard (Payments)}& {\tiny 10.6 
 }& {\tiny 2.8 }& {\tiny 16.3 }& {\tiny 97\% 
 }\\ 
 {\tiny Sending messages (Lack of Guidance)}& {\tiny 2.4 
 }& {\tiny 1.9 }& {\tiny 0.7 }& {\tiny 51\% 
 }\\
 \rowcolor{white}
 {\tiny Watching over requesters' profiles (Hypervigilance)}& {\tiny 15.0 
 }& {\tiny 1.1 }& {\tiny 12.9 }& {\tiny 69\% 
 }\\
 {\tiny Searching for general HITs (Hypervigilance)}& {\tiny 3.6 
 }& {\tiny 0.9 }& {\tiny 5.6 }& {\tiny 96\% 
 }\\
 \rowcolor{white}
 {\tiny Managing queued HITs (Hypervigilance)}& {\tiny 3.2 
 }& {\tiny 0.7 }& {\tiny 4.6 }& {\tiny 93\% 
 }\\
 {\tiny Previewing HITs (Lack of Guidance)}& {\tiny 1.5 
 }& {\tiny 0.6 }& {\tiny 1.0 }& {\tiny 66\% 
 }\\
\rowcolor{white}
 {\tiny Viewing their earnings (Payments)}& {\tiny 0.9 
 }& {\tiny 0.5 }& {\tiny 0.3 }& {\tiny 85\% 
 }\\
 {\tiny Searching for filtered HITs (Hypervigilance)}& {\tiny 3.9 
 }& {\tiny 0.5 }& {\tiny 0.6 }& {\tiny 46\% 
 }\\
 \rowcolor{white}
 {\tiny Checking Worker's qualifications (Hypervigilance)}& {\tiny 0.4 
 }& {\tiny 0.2 }& {\tiny 0.0 }& {\tiny 27\% 
 }\\
 {\tiny Login to MTurk (General Logistics)}& {\tiny 0.3 
 }& {\tiny 0.1 }& {\tiny 0.1 }& {\tiny 64\% 
 }\\
 \rowcolor{white}
 {\tiny Reading HIT information (Lack of Guidance)}& {\tiny 0.1 
 } & {\tiny 0.0 }& {\tiny 0.0 }& {\tiny 63\% 
 }\\
 {\tiny Reading Platform Help (Lack of Guidance)}& {\tiny 0.0 
 }& {\tiny 0.0}& {\tiny 0.0}& {\tiny 0\% 
 }\\

\end{tabular}
\caption[Table caption text]{Overview of the invisible labor activities that workers did, the amount of time they dedicated daily to each activity per day, and the percentage of workers who engaged in the activity. Doing HITs that eventually time out was the median most time consuming activity; viewing their earnings was the most common activity.}
\label{table:invisibleLaborTypes}
\end{center}
\end{table}

\begin{table}[t]
\begin{center}
\begin{tabular}{c c c c }
 {\bf \tiny Main Category of Invisible Labor} & {\bf \tiny Mean}& {\bf \tiny Median}& {\bf \tiny Std}\\ 
 \hline
 \rowcolor{white}
 {\tiny Payments}&{\tiny 14 
 min}&{\tiny 13 min}&{\tiny 23.8}\\
  {\tiny Hypervigilance}&{\tiny 28 
 min}&{\tiny 11 min}&{\tiny 56.8}\\
 \rowcolor{white}
 {\tiny Lack of Guidance}&{\tiny 16 
 min}&{\tiny 6 min}&{\tiny 62.1}\\
 {\tiny Breaks}&{\tiny 3 
 min}&{\tiny 3 min}&{\tiny 12.6}\\
 \rowcolor{white}
 {\tiny General Logistics}&{\tiny 1 
 min}&{\tiny 1 min}&{\tiny 0.1}\\
 \end{tabular}
\caption[Table caption text]{Overview of the categories of invisible labor that workers did and the median amount of time they dedicated to it daily. The category of Payments was the one workers invested the most median time daily.}
\label{table:invisibleLaborCategory}
\end{center}
\end{table}

\subsubsection{Quantifying Categories of Invisible Labor.} We were also interested in understanding the type of invisible labor that was the most taxing for workers. Table \ref{table:invisibleLaborTypes} presents an overview of the different invisible labor activities that our plugin detected that workers did and the percentage of workers who engaged in each activity. For each activity, we also present in parenthesis the main categories to which the activity belongs. In Table \ref{table:invisibleLaborCategory}, we present a summary of the time workers invested in each of these main categories. From Tables \ref{table:invisibleLaborTypes} and \ref{table:invisibleLaborCategory}, we observe that the invisible labor category of  ``Payments'' was the most time-consuming category (especially when taking the median value) and was also highly common among workers. For example, Table \ref{table:invisibleLaborTypes} shows how 97\% of all workers in our study engaged in the Payments related activity of checking their daily earnings on their worker dashboard. Similarly, the most time-consuming activity was ``doing HITs that eventually timeout,'' which took a median of 4.5 minutes. Luckily, timeouts were not as common (only 37\% of workers engaged in this activity). It is important to mention that timeouts relate to ``Payments'' because requesters on MTurk have to specify the amount of time that workers have for completing their  tasks; if workers take longer than that time, the HIT is timed out, and workers do not get paid for any of the labor that they did for the HIT.
We calculated timeouts only if the worker was actually working on the HIT (had any current mouse or keyboard-related activity on the HIT). The timeouts we detected were, therefore, cases where the worker was actively doing labor but at the end did not get paid for it. 

To understand the details of the workers who engaged the most in this type of highly taxing invisible labor, we first identified the workers who were outliers (i.e., invested the most time in this activity) and then conducted a manual inspection of their digital traces. We considered outliers to be the workers whose time invested for this particular activity was above the 95th percentile (typical method to calculate outliers \cite{hao2007quantile}). We observed that in this case, the outliers tended to be workers who accepted a high number of HITs within a given time window (likely to avoid having other workers take the HIT before them). However, the problem was that it would sometimes  take workers significant time to get to some of the HITs they had ``reserved'' for themselves, and hence they experienced timeouts. We thus believe there is value in exploring tools \cite{lioznova2020prediction}, that based on workers' log data, can automatically learn the best amount of time that should be allocated for a given task and then recommend to requesters to use a significantly higher time window than that time to avoid timeouts and also be sympathetic with the labor practices of some workers.

Table \ref{table:invisibleLaborCategory} also shows that the second most time-consuming category was that of Hypervigilance, taking workers' a median of 11 minutes daily. The Hypervigilance activity that took the most time was watching over requesters' profiles. It is likely that workers engaged in this activity because through this they could more easily grab the HITs that their favorite requesters posted \cite{gray2019ghost}. Upon manual inspection of workers' digital traces, we identified that the workers who invested the most time in this activity (i.e., the outliers, which we calculated with a similar method as stated above), were the workers who appeared to hunt the profiles of multiple requesters ready to be ``on-call''. (In specific, these workers opened the profile pages of multiple requesters and then iterated through the list of profile pages, likely inspecting if the requesters had posted anything new.)

Finally, the third most time-consuming {\it category} was ``Lack of Guidance,'' which took a median of six minutes daily. The most time-consuming activity here were cases when workers started a HIT but then decided to return it. There are several reasons why workers might engage in this behavior; for example: workers realize that the HIT is more time-consuming than they expected; or the HIT involves skills that the worker lacks; or the HIT consists of activities that the worker does not enjoy. In general, these are instances where the HIT instructions likely did not correctly guide the worker on the type of labor to expect, and hence the worker had to return the HIT. Prior work has already reported how the lack of guidance can lead to these types of dynamics \cite{manam2018wingit, gadiraju2017clarity}. From Table \ref{table:invisibleLaborTypes}, we note that the activities related to the Lack of Guidance were actually some of the most commonplace for workers and also some of the most time consuming (e.g., 
92\% engaged in starting HITs but then returning them; and this was also the second most time consuming activity.) It was surprising to see the large percentage who returned HITs. Upon manual inspection of the outliers, we observed that they appeared to primarily follow a discard-by-doing labor pattern \cite{kelley2010conducting}.

\begin{table}[t]
\centering 
\begin{tabular}{c c c c} 
\hline 
{\tiny Perception of Time in Invisible Labor} & {\tiny Percentage of Workers} & {\tiny Perceived Time}& {\tiny Actual Time} \\ [0.5ex] 
\hline 
{\tiny Far too much time} & {\tiny 25\%} & {\tiny 3 hrs 4 min} & {\tiny 2 hrs 23 min} \\ 
{\tiny Too much time} & {\tiny 38\%} & {\tiny 2 hrs 3 min} & {\tiny 1 hr 40 min}\\
{\tiny An adequate amount of time} & {\tiny  29\%} & {\tiny 1 hr 50 min} & {\tiny 1 hr 32 min} \\
{\tiny Too little time} & {\tiny 5\%} & {\tiny 1 hr 15 min} & {\tiny 56 min} \\
{\tiny Far too little time} & {\tiny  3\%} & {\tiny 1 hr} & {\tiny 40 min}\\  
\hline 
\end{tabular}
\caption{Summary statistics of workers' perceptions of how much time they felt they invested in invisible labor. Notice that the perceived and actual times are the medians for each perception group.} 
\label{table:p}
\end{table}

\begin{figure}
  \begin{center}
    \includegraphics[width=.6\linewidth]{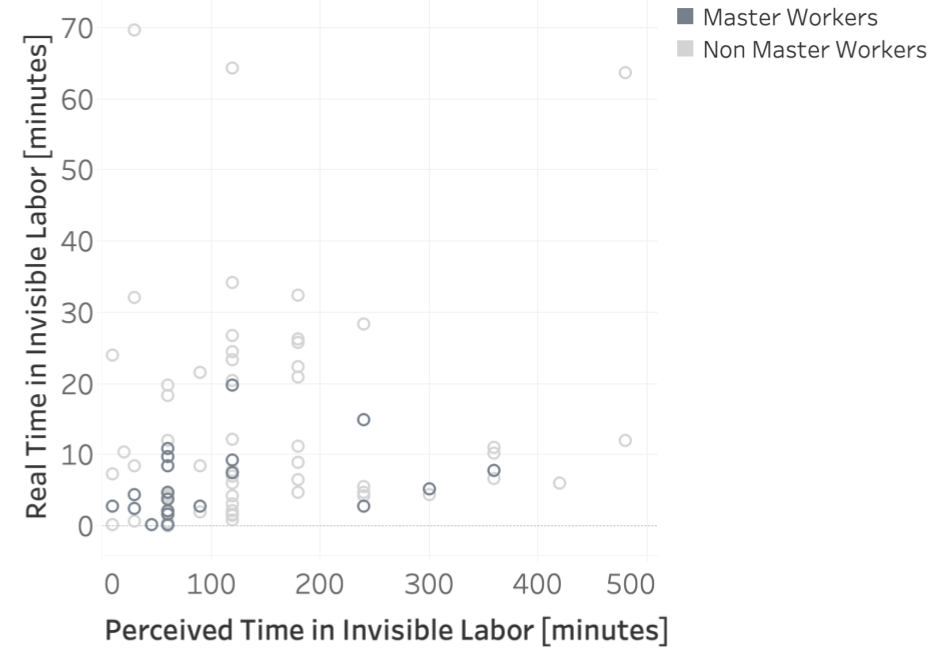}
  \end{center}
  \caption{Comparisons of the estimated and actual time that workers invested in invisible labor. Most workers overestimated how much time they dedicated to invisible labor.}
  \label{fig:expectedLabor}
  \vspace{-4pt}
\end{figure}

\subsection{Perceptions of Invisible Labor}
\label{sec:perceptions}
Workers from digital labor platforms typically underestimate the actual amount of time and effort they dedicate to invisible labor \cite{whiting2020digi}.
However, workers' perceptions of invisible labor can play a strong role in how they feel about their work. In this section, we investigate the amount of time that crowd workers believe they invested in invisible labor and their satisfaction. For this part, we use the initial survey that we gave workers, which was inspired by prior work \cite{bernstein2013quantifying}. Through this, we found that workers in our study estimated that they spent a median of 2 hours daily on invisible labor on MTurk (with the minimum time that some workers' estimated as 0 and a maximum of 8 hours.) Figure \ref{fig:expectedLabor} plots the actual time workers invested in invisible labor against perceived time.  Notice that each point represents a worker in our study, and workers are color-coded based on whether workers are master-workers (dark gray) or non-master workers (light gray). We made this distinction given that prior work has identified that there are differences in how more experienced workers operate \cite{han2020crowd,savage2020becoming}, and our results in the previous sections were also highlighting these differences. Notice that workers in Figure \ref{fig:expectedLabor} who were able to accurately guess the amount of time they spent on invisible labor are located on the diagonal line, as that is when the actual time is equal to the perceived time. The cluster of points that we observe above the diagonal line close to the Y-axis showcases that the majority of crowd workers in our study overestimated the amount of time that they thought they invested in invisible labor. Notice that this overestimation occurred for both master and non-master workers. Next, we quantify the relative error of workers in estimating how much time they invested daily in invisible labor:
\begin{equation}
    1-\left(\frac{estimated\_invisible\_labor\_time}{actual\_invisible\_labor\_time}\right)
\end{equation}

Through this, we identified that the median relative error was $-0.14$. Notice that the negative value highlights that workers are overestimating how much time they dedicate to invisible labor, but our conservative metrics used to quantify invisible labor might also contribute to the error. Next, we analyze workers' satisfaction with the time they perceived they invested in invisible labor. Table \ref{table:p} presents a summary of these statistics. Only 10\% felt they invested too little time on invisible labor (8\% felt they spent ``too little time'' and 2\% ``far too little''), while slightly more than half (63\%) felt they invested too much in invisible labor (38\% ``too much time,'' and 25\% ``far too much time''). Lastly, 27\% of workers considered they invested the right amount of time in invisible labor. Future work could study the type of labor dynamics that might lead workers to feel more satisfied with the amount of invisible labor that they do, and also what circumstances might lead them to feel the most dissatisfied.

\section{Discussion}
The core result from our study is that crowd workers spent a median of 33 minutes 
of their daily time on MTurk doing invisible labor, and this labor leads workers to drop their median hourly wage from \$3.76 to \$2.83. Notice that because we used conservative methods to measure invisible labor, we are obtaining a lower bound of the amount of invisible labor that exists on MTurk. However, this lower bound is still highlighting and providing  quantitative support to the literature's qualitative claim that invisible work makes up a substantial fraction of the work done in crowdsourcing markets and, therefore, dramatically reduces workers' hourly wages \cite{gray2019ghost}. Considering that the median hourly wage of workers is just  \$3.76 (without considering invisible labor), it is clear that crowd workers still need a dramatic increase in their wages before we can consider this labor fair. However, this is not only something for requesters to consider, but also something for platforms, workers, and even policy makers. In this section, we discuss: the details of the most taxing categories of invisible labor that our study uncovered; design and policy solutions to mitigate invisible labor on crowdsourcing platforms. Additionally, we make an effort to connect with invisible labor in other workplaces, as well as with critical theory, to have a broader discussion on the implications of our research.

\subsubsection{Most Common and Most Time-Consuming Invisible Labor.} The invisible labor that the overwhelming majority of workers in our study practiced was around Payments. In fact, 97\%  
of the workers in our study visited the earnings section on their worker's dashboard at least once daily. Crowd workers are likely visiting their earnings dashboard to ensure that they: (1) were paid for their labor; and (2) made a certain daily income amount \cite{kaplan2018striving}. For the first point, it is important to note that crowd workers typically have to deal with faceless requesters, machines that are outdated, unreliable internet connections, and have nowhere to report when things go wrong (e.g., report that a requester decided to unjustly withhold payment, or report that due to technical issues they can no longer access their MTurk account and earnings.) Pew Research reported that 30\% of on-demand gig workers experienced situations where they were not paid for their labor \cite{hitlin2016research,smith2016labor}. Similarly, the US Freelancers Union found that 71\% of freelancers have struggled to collect payment for their work. As we note, ensuring payment is a critical and stressful aspect of crowd work \cite{costs_2019}. For the second point, we have to be aware that most crowd workers struggle to make a minimum wage \cite{hara2018data}. Therefore, another likely reason why workers were visiting their earnings dashboard was to see if they had made sufficient wages. The stress of not receiving payment for their labor or not receiving enough appears to be very present and real in crowd work. 

Invisible labor around payments was actually also the most time-consuming, and one of the most critical, as it relates to workers' livelihood. To address this problem, designers could explore interfaces where workers are constantly informed of their current earnings. However, seeing their earnings constantly could also create stress on workers. Future work could explore optimal settings for displaying wages in crowd work. We also believe there is value in further exploring interfaces where requesters, platforms, and workers agree to fair wages \cite{silberman2018responsible}. Offering workers a space where they know they will be treated fairly could reduce repeated payment checking \cite{whiting2019fair}. 

It is important to mention that this type of invisible labor is also present in other digital workspaces \cite{rosenblat2018uberland,hall2015effects}. For example, Uber had reports of drivers and passengers organizing to check how much a passenger was actually charged for a ride vs. how much the driver received. This dynamic emerged after Uber changed its pricing algorithm and did not provide transparency on how it functioned \cite{chen2015peeking}. The lack of transparency not only led drivers and passengers to have to engage in this type of invisible labor, it also led them to feel cheated and betrayed by the platform \cite{rosenblat2018uberland}. Here it is important to highlight that this invisible labor does not only emerge due to the fault of requesters (passengers). But rather, platforms can play a key part in the promotion of this invisible labor. Here it can be important for platforms to see that this type of invisible labor is likely emerging out of mistrust and has the potential to alienate people from their platforms.

\subsubsection{Second Most Time-Consuming Category of Invisible Labor.}Our study uncovered that the invisible labor category of
Hypervigilance was the second most time-consuming for crowd workers. Workers spent a median of 11  minutes daily on this category of invisible labor. Crowd work has been championed as offering people the unique flexibility of working anytime and from anywhere \cite{yin2018running,deng2013crowdsourcing}. However, our work highlights how this flexibility is likely more of a myth. Crowd workers have to dedicate significant time daily to search for work and be on-call for requesters. Intuitively, this suggests that there are more workers on the site than there is work to be done. (If there were lots of requesters constantly posting lots of high-paying jobs, workers would not feel the need to be on call to get the good work.)
This connects with prior work that shows that requesters have the majority of the power in this market partly due to the fact that there is an extreme concentration of a few requesters who post the majority of the tasks \cite{dube2018monopsony,kingsley2015friction}. Thus, workers are forced to take whatever jobs at whatever pay these few requesters post. 

A way to start addressing this problem could be to build off the different tools and computational methods that have been developed to achieve fair compensation \cite{whiting2019fair}. Potentially these computational methods could be extended by incorporating an invisible labor component. For instance, workers could be computationally guided to cooperate with each other to ensure fairer wages and minimize the amount of invisible labor in which they engage \cite{fan2020crowdco}, such approach could be extended to potentially lead to reduced invisible labor. Similarly, we could also consider how algorithms that facilitate automatic task assignment and recommendations \cite{hettiachchi2020crowdcog}, could be helpful in reducing invisible labor by minimizing the task search time.

When thinking about the invisible labor around Hypervigilance, it is also important to notice that this type of invisible labor is one that promises workers high returns (especially as by being vigilant, workers can potentially earn high wages). 
Here, it can be important to identify that other digital labor platforms have started to weaponize this type of invisible labor to manipulate workers to stay longer on their platforms \cite{guda2019your, wang2019ridesourcing}. For example, Uber sends drivers messages to motivate them to keep being vigilant of surge pricing \cite{chen2015peeking}. The following is an example message that Uber sends drivers to motivate them to remain vigilant of surge pricing: \textit{``The weekend is here, and demand is on the rise in Lehigh Valley! Plan to go online tonight, and keep an eye out for surge around the area, where you can earn over 3X on fares! Stay online through midnight to take advantage of the highest fares. Uber on!''} \cite{rosenblat2018uberland}. In this context, we believe there is value in providing workers with tools that can help them to visualize how digital labor platforms might be manipulating them to engage them in free labor. Related, there is likely also value in tools that can inform workers of the likelihood of achieving specific wages if they engage in hypervigilance within particular time windows.  

\subsubsection{Invisible Labor in Other Workplaces and Policy.}
Researchers have argued that within our ``capitalist societies'', there is a propensity to manage the workforce in ways that will profit the ``capitalists'' (who in this context could be considered to be Amazon or the requesters.) \cite{federici1975wages, d2020data}. Such ``workforce management'' can include defining  what labor is counted and what labor is turned invisible \cite{davis1983approaching,daniels1987invisible}. Labor visibility (what is counted) is considered to be especially important in this societal context because the cultural worth of a piece of labor is directly connected to how much the labor costs \cite{moulier2011cognitive}. Work that is done for free (invisible) usually will fail to be valued \cite{beechey1977some,federici1975wages, terranova2000free}. Several labor collectives, researchers, practitioners, and individual citizens have therefore fought to empower workers to gain visibility and recognition for their work \cite{frazis2012think, d2020data}. For instance, the International Feminist Collective has been fighting for decades to give more visibility to the housework that women perform \cite{federici1975wages}. The collective has argued that housework  has been undervalued, underpaid, and its invisibility has been used as a means to empower primarily ``white middle-class men to do lucrative waged jobs,'' e.g., office work \cite{davis1983approaching, folbre2007child}. This in return has profited companies and factories as they now have a more specialized and dedicated workforce \cite{lim1983capitalism, floro2011gender,floro2010gender}).

In 2013, several of these collectives had a breakthrough when labor statisticians agreed internationally to begin measuring in official workforce surveys both paid and unpaid labor, such as housework \cite{buvinic2018invisible, cepal2015classification}. This inclusion influenced the development of new policies around invisible labor \cite{cepal2015classification, buvinic2018invisible}. Historically, policymakers had overlooked unpaid labor simply because the work was not included in the official statistics that they used to define policy \cite{weyrauch2011sound, buvinic2018invisible}. Its exclusion also meant that policymakers did not understand why the labor was problematic or the number of citizens who were impacted. But, by now counting and including the labor within the official stats that policymakers used for their decision-making, they were able to more easily pay attention to this type of labor, grasp its problematic, and design policy to address the challenges.

Inspired by the impact that the quantification of invisible labor has had in transforming policy within other industries and workplaces, our hope is that our plugin tool, study, and anonymized worker data, can in the future also be used to motivate new policies to improve the labor conditions of crowdworkers. However, given that the use of data in policymaking is usually an organic, political process \cite{dhaliwal2012research} (which might not be obvious to outsiders, e.g., workers and their advocates), we believe there is value in designing socio-technical mechanisms that guide citizens on how they can best use the data from our plugin to drive policy innovation \cite{crewe2002bridging}. This could include tools that guide citizens on the time in which they should release the data on invisible labor to match the political cycle. Being in tune with the political cycle could help citizens to have a better chance at influencing policymakers \cite{weyrauch2011sound}. Similarly, other tools could focus on helping citizens to easily visualize which policymakers might be most influenced by seeing the stats from our plugin on invisible labor. There is likely also value in tools that can guide citizens on how to use our plugin's data to gather the public's support and create pressure on policymakers \cite{blagescu2006capacity, start2004tools}.

\subsubsection{Design Implications \& Future Work.} 
Future work could explore mechanisms to help workers manage the time overheads from invisible labor. Notice that here there are still numerous aspects of invisible labor that need to be further investigated. For instance, are more experienced crowd workers able to reduce the amount of time they spent in invisible labor in comparison with novices? Our results highlight that at least master and non-master workers have similar perceptions of the amount of invisible labor they do. But more analysis in this space is necessary. Especially because there might be a benefit in designing tools that help novice workers adopt some of the strategies from more experienced workers \cite{han2020crowd,savage2020becoming}. Other questions we are interested in exploring in this space are: How does the way that workers manage their invisible labor relate to their wages? How exactly does multi-tasking and context switching relate to invisible labor? Is a worker's invisible labor increased when workers have to switch between HITs? Are there certain HITs or requesters that magnify workers' invisible labor?  
Our hope is that by releasing our plugin, we will enable the scientific community to study this.

Notice that our plugin tool can be easily extrapolated to other digital labor platforms to help workers quantify  the amount of invisible labor that they spend on those other workplaces (the only main piece that needs to be changed is the mapping between the websites the workers use and the work done on each platform; primarily if it is paid or unpaid labor). Our hope is that our tool will inspire cross-platform studies on invisible labor and will help the scientific community to derive principles around how invisible labor looks like across digital workplaces. As we described above, our plugin and study could also help to motivate action from policymakers. 
Facilitating tools for cross-platform auditing can be extremely important as digital labor platforms have traditionally been black boxes. But, to design better platforms or drive policy change, it is crucial to understand what happens inside these platforms. Our hope is that our research will be a step forward to better understand and address the dynamics existing in these online spaces. 

We believe there is likely value in exploring data visualizations that could help to better showcase the different types of invisible labor that crowd workers have to do. Here, we could take inspiration from the visualizations that Github has developed to showcase the labor surrounding the writing of collaborative code \cite{d2020data,liao2018exploring}. Github has made great strides to provide visualizations that help people to rapidly understand the quantity, frequency, and duration of the contributions made by each individual to a codebase. Such visualizations in this context could help requesters to better grasp the amount and type of invisible labor that their tasks are forcing workers to do and potentially lead requesters to better compensate workers for their effort and time \cite{whiting2019fair}. It is important here to consider how to design such visualizations to also not incite unhealthy competition between workers or enable abuse and surveillance from requesters \cite{kittur2012crowdweaver,rzeszotarski2012crowdscape}. 

\subsubsection{Critical Theory and Design to Address Invisible Labor in Different Digital Workplaces.} 
An important question in CSCW is whether a new design truly engages with the root cause of a societal problem or if it is primarily dealing with the symptoms of a problem \cite{bardzell2018critical}. For example, a design could make a societal problem bearable. However,  this might lead people to no longer have a need for addressing the root problem. In this setting, the design could provide enjoyable experiences to end-users; but it could also reinforce the structural issues that are harming end-users. Within this context, Herbert Marcuse, a theorist from the Frankfurt School of Critical Theory \cite{jay1996dialectical}, introduced the concept of ``one-dimensional'' people who have a conformist understanding of society that does not allow them to critique or question how society could be different \cite{marcuse2013one}. Marcuse argues that the one-dimensional person has lost her ability to critique society because consumerism has tricked her into having false needs and wishes (notice that consumerism is considered to be ``a social and economic order that encourages the acquisition of goods and services in ever-increasing amounts'' \cite{britannica2008britannica}). As a result, the person focuses on fulfilling those ``fake needs'' instead of questioning the problematic societal structures in which she is immersed. According to Marcuse, this dynamic leads us to be imprisoned into one-dimensional thinking, and that makes it extremely challenging to critically question the structures and processes that exist in our society. 

As CSCW researchers, we believe it is crucial that we question to what extent we are falling into one-dimensional thinking and possibly strengthening the structural issues that are already in place. This is especially important when designing interfaces that aim to address the problem of invisible labor in crowdwork and also within other digital labor platforms. Without this critical analysis, we might fall into designing interfaces that make the problem of invisible labor bearable; but we never address the systematic problems surrounding workers, requesters, and digital platform owners. Notice that engaging in such critical analysis is an ambitious, complex, and difficult undertaking, but as Marcuse discusses, it is very much necessary \cite{bardzell2018critical,marcuse2013one}. 

Marcuse argues that a way to engage in such critical analysis and challenge our one-dimensional thought is by participating in artistic creativity that allows us to leave the reality that has been defined by society \cite{marcuse2013one, marcuse1966individual, bardzell2018critical}. Artistic creativity facilitates developing new designs that are not confined by the current reality of what is possible and allows us to consider designs we might have been blind to consider otherwise. Based on this, we believe there is value in engaging with workers in ``creative artistic co-design sessions.'' These sessions would allow workers to creatively define the type of digital labor platforms that they would like to see and how they would design to address invisible labor \cite{toxtli2020meta}. Similarly, we believe there is value in drawing on scholarship that has studied the link between fiction and design \cite{wong2017real,lindley2015back,disalvo2016designing,fiesler2021innovating}. Here we envision we could engage researchers, workers, platform owners, and practitioners to use fictional narratives to design ``alternative realities'' to contemporary digital labor platforms and tools \cite{linehan2014alternate,fiesler2018owning}.

\subsubsection{Limitations.} The insights from our research are limited by the methodology and population we studied. Our study also focused on breadth instead of depth to start to shed needed light on the quantification of invisible labor in crowd work. Notice that we had to develop specific tools in order to do our field study, which is not simple. However, these types of studies are important, especially given the lack of transparency that MTurk or other crowdsourcing platforms provide around invisible labor. Upon publication, we will open-source our plugin and anonymous worker data so that the scientific community can conduct longitudinal studies around invisible labor, as well as study other principles surrounding invisible labor.

\section{Conclusion}
We developed a new computational tool to be able to quantify and study the invisible labor of crowd workers on MTurk. We have demonstrated that the invisible labor that workers do can take a toll on their wages. Particularly, we saw that if we consider the amount of time that crowd workers invest in invisible labor, their hourly wages go down to \$2.83 from \$3.76.  We also identified that the two most time-consuming categories of invisible labor revolved around payments and hyper-vigilance. Additionally, our study identified that workers tended to overestimate the amount of invisible labor that they believed they did.  
  Our results also suggest there is a wide range of dynamics that influence the amount of invisible labor that a particular worker conducts. These different dynamics deserve more investigation.
  
Finally, we hope that our plugin tool inspires the auditing of different digital labor platforms and helps to potentially generate a range of positive policy innovations in digital work. Our paper has provided much-needed light to the invisible labor of crowd workers.  

{\bf Acknowledgments}. Special thanks to all the anonymous reviewers who helped us to strengthen the paper. This work was partially supported by NSF grant FW-HTF-19541.

\bibliographystyle{ACM-Reference-Format}
\bibliography{main}


\end{document}